\documentclass{PoS}
\pdfoutput=1

\title{Jet substructure measurements in pp collisions at $\sqrt{s}$ = 13 TeV with ALICE}

\ShortTitle{Jet substructure measurements in pp collisions at $\sqrt{s}$ = 13 TeV with ALICE}

\author{\speaker{Markus Fasel for the ALICE Collaboration}\\
        Oak Ridge National Laboratory, Oak Ridge, Tennessee, United States\\
        E-mail: \email{markus.fasel@cern.ch}}

\abstract{ We present a variety of jet substructure measurements performed in pp collisions at 
$\sqrt{s}$ = 13 TeV with the focus on the groomed jet momentum fraction $z_{\rm{g}}$ in a wide range
of $p_{\rm{T}}$ between 20 and 200 GeV/$c$ and jet resolution $0.2 < R < 0.5$. Thanks to the capabilities 
of the ALICE apparatus jet substructure measurement are possible with an infrared constituent cutoff 
at 0.3 GeV. Furthermore, the angular resolution of the ALICE detectors allows the measurement of jet 
substructure observables with a high precision. The measurements are compared to pQCD calculations and MC generators. Furthermore, the measurement of track-based jets at the same centre-of-mass energy and its dependence on the event activity are presented for different jet resolutions.}

\FullConference{International Conference on Hard and Electromagnetic Probes of High-Energy Nuclear Collisions\\
		30 September - 5 October 2018\\
		Aix-Les-Bains, Savoie, France}

\begin{document}

\section{Introduction}

Jet substructure measurements are an ideal tool to constrain the perturbative and non-perturbative effects - consisting of perturbative radiation, hadronization and contributions from the underlying event - contributing to the jet shower. Particularly at low-$p_{\rm{T}}$ where the effects can become large the measurements at different jet resolution parameter become effective in disentangling the contributions from the various effects \cite{Dasgupta:2016bnd}. In addition, the increased statistics taken in the new datasets allow for more differential measurements. This is exploited in measurements of the groomed momentum fraction applying the SoftDrop algorithm \cite{Larkoski:2014wba,Butterworth:2008iy}

\begin{equation}
    z_{\rm{g}} = \frac{min(p_{\rm{T},1}, p_{\rm{T}, 2})} {p_{\rm{T},1} + p_{\rm{T},2}} > z_{cut} \theta^{\beta}
\end{equation}

with $p_{\rm{T},1}$ and $p_{\rm{T},2}$ being the $p_{\rm{T}}$ of the hardest and second hardest subjet. The level of grooming is steered by the parameters $z_{cut}$ and $\beta$. The observable is selected as it closely relates to the QCD splitting function at leading order \cite{Dasgupta:2016bnd}. As long as this connection holds, we should not observe a jet momentum dependence of $z_{\rm{g}}$, which we will test. At leading order, the difference between the parton and the reconstructed jet momentum is expected to increase as log($R$) with perturbative radiation, to decrease as 1/$R$ due to hadronization effects and to increase as $\sim R^{2}$ due to the underlying event \cite{Dasgupta:2016bnd}.  The $z_{\rm{g}}$ and jet substructure in general allow for more differential inspection of the previous components.

The data used for the measurements were collected in 2016 and 2017. It consists of a min. bias dataset corresponding to an integrated luminosity $L_{\rm{int}} = 11.5~\rm{nb}^{-1}$. Measurements of fully reconstructed jets also utilize a $L_{\rm{int}} = 4~\rm{pb}^{-1}$ dataset collected with the jet trigger implemented in the electromagnetic calorimeter for jets with $p_{\rm{T}} > 100~\rm{GeV}/c$.

Jets are reconstructed using the FastJet package \cite{Cacciari:2011ma} applying the anti-$k_{T}$ algorithm \cite{Cacciari:2008gp} with the E-scheme for recombination. Track-based jets are built only of tracks reconstructed in the central tracking detectors with a minimum track $p_{\rm{T}}$ of 0.15 $\rm{GeV}/c$. For fully reconstructed jets also energy carried by neutral particles measured in the electromagnetic calorimeter is included. Neutral constituents must have a minimum energy of 0.3 GeV. The average shift of the jet energy scale is $\sim15\%$ for track-based jets, compared to particle-level jets consisting only of charged particles, and $\sim25\%$ for fully reconstructed jets with $p_{\rm{T}} \sim 40~\rm{GeV}/c$.

\section{Charged jet production as function of the event activity}

\begin{figure}
    \centering
    \includegraphics[width=0.3\textwidth]{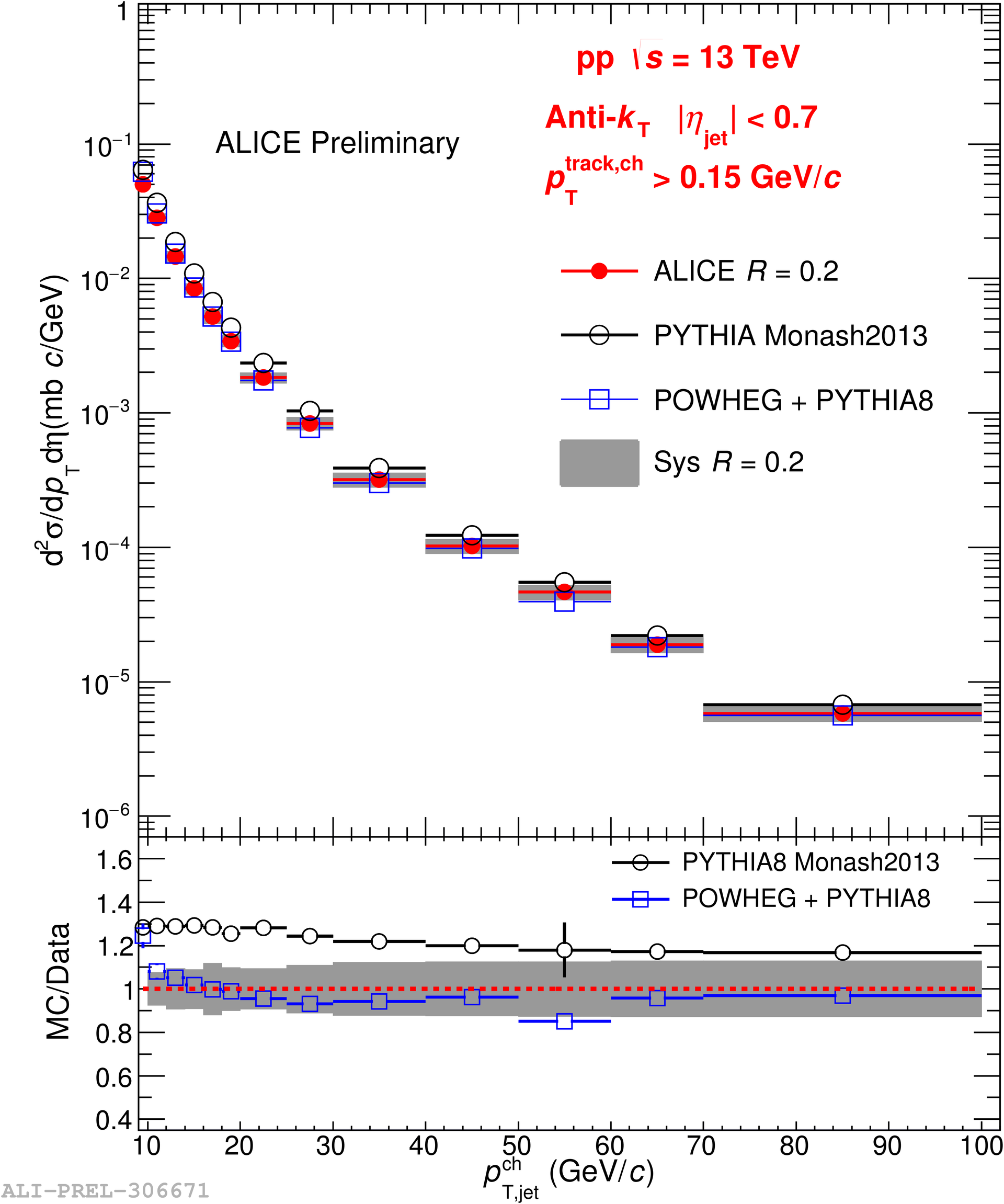}
    \includegraphics[width=0.3\textwidth]{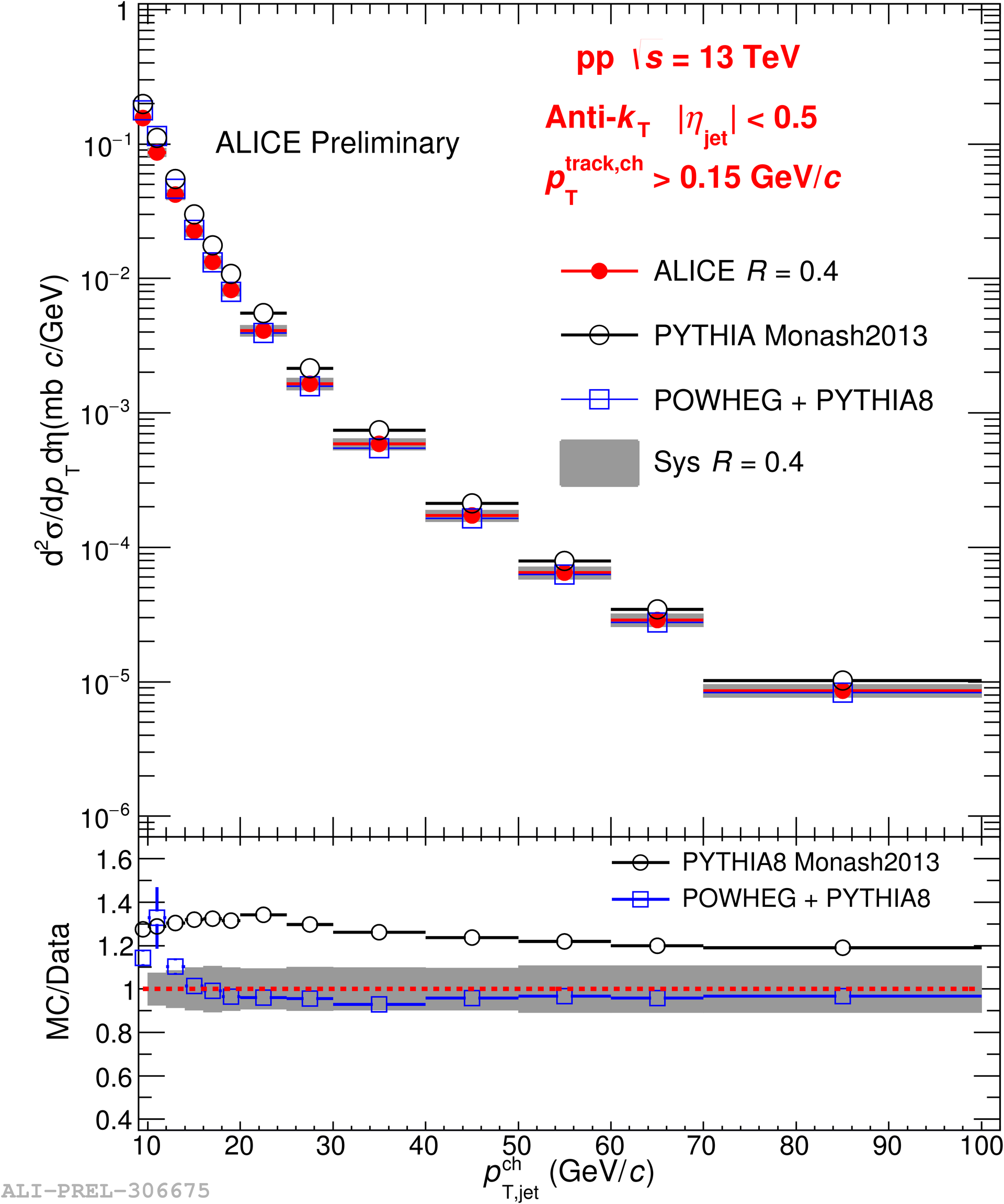}
    \includegraphics[width=0.3\textwidth]{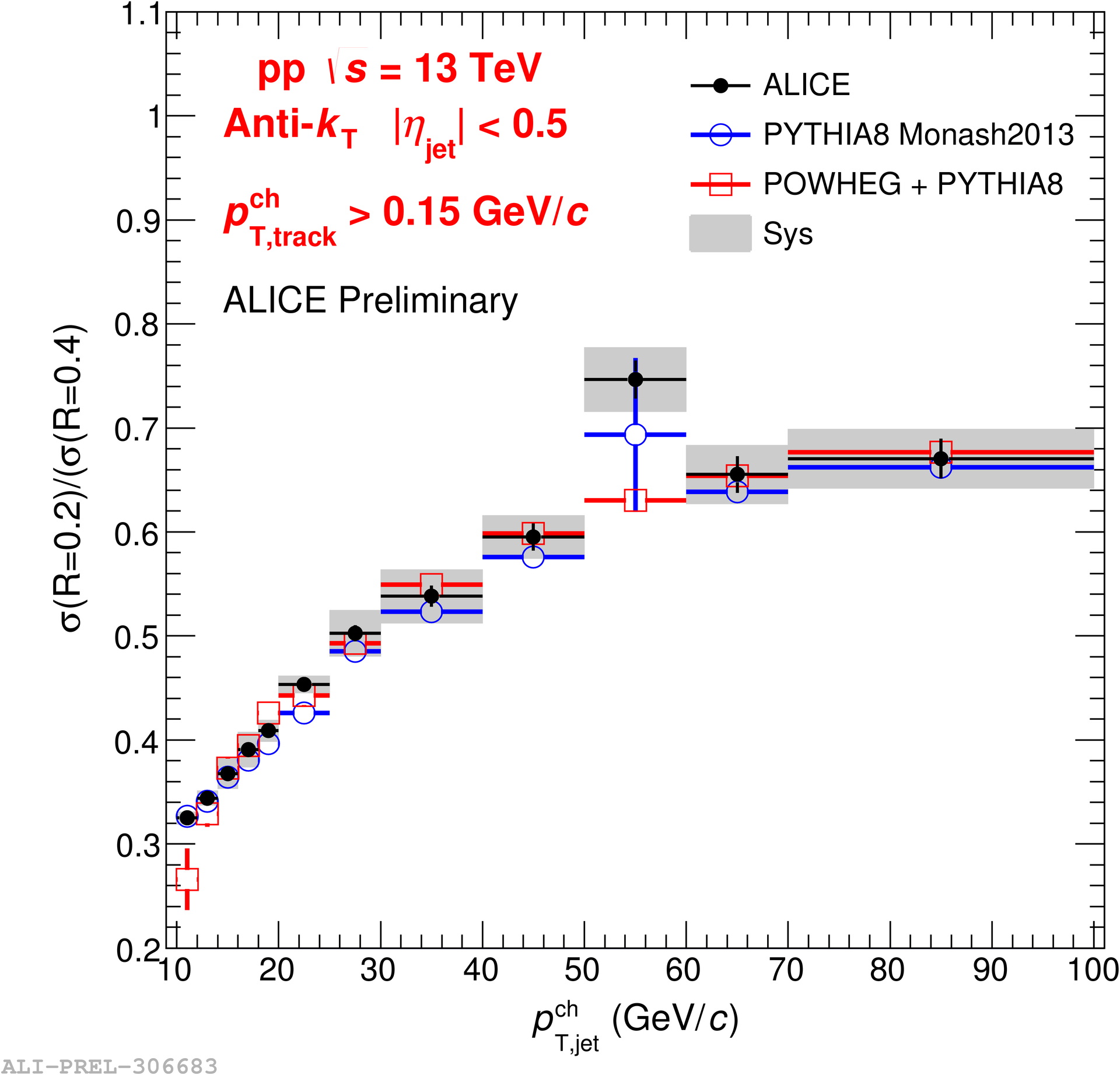}
    \caption{$p_{\rm{T}}$-differential cross section of the production of track-based jets for $R$=0.2 (left) and $R$=0.4 (middle) compared to the PYTHIA8 event generator using the Monash2013 tune and POWHEG+PYTHIA. Lower panels show the ratio MC to data. Right: Cross section ratio of track-based jets with $R$=0.2/$R$=0.4 compared to PYTHIA8 Monash 2013 and POWHEG+PYTHIA.}
    \label{fig:chargedjets13TeV}
\end{figure}

The $p_{\rm{T}}$-differential cross section of the production of track-based jets, subtracted for the contribution from the underlying event, is shown in Fig.~\ref{fig:chargedjets13TeV} for $R$=0.2 (left) and $R$=0.4 (middle). The spectra are compared to calculations utilizing the PYTHIA8 \cite{Sjostrand:2014zea} event generator using the Monash2013 tune and POWHEG \cite{Nason:2004rx, Frixione:2007vw, Alioli:2010xd, Alioli:2010xa} next-to-leading order event generator interfaced with PYTHIA8. While PYTHIA8 overpredicts the cross section by $~20\%$ calculations using POWHEG are in good agreement with the the data. The cross section ratio, shown on the right side, increases with jet $p_{\rm{T}}$ as jets become more collimated. The trends are well described by both PYTHIA8 and POWHEG.

\begin{figure}
    \centering
    \includegraphics[width=0.3\textwidth]{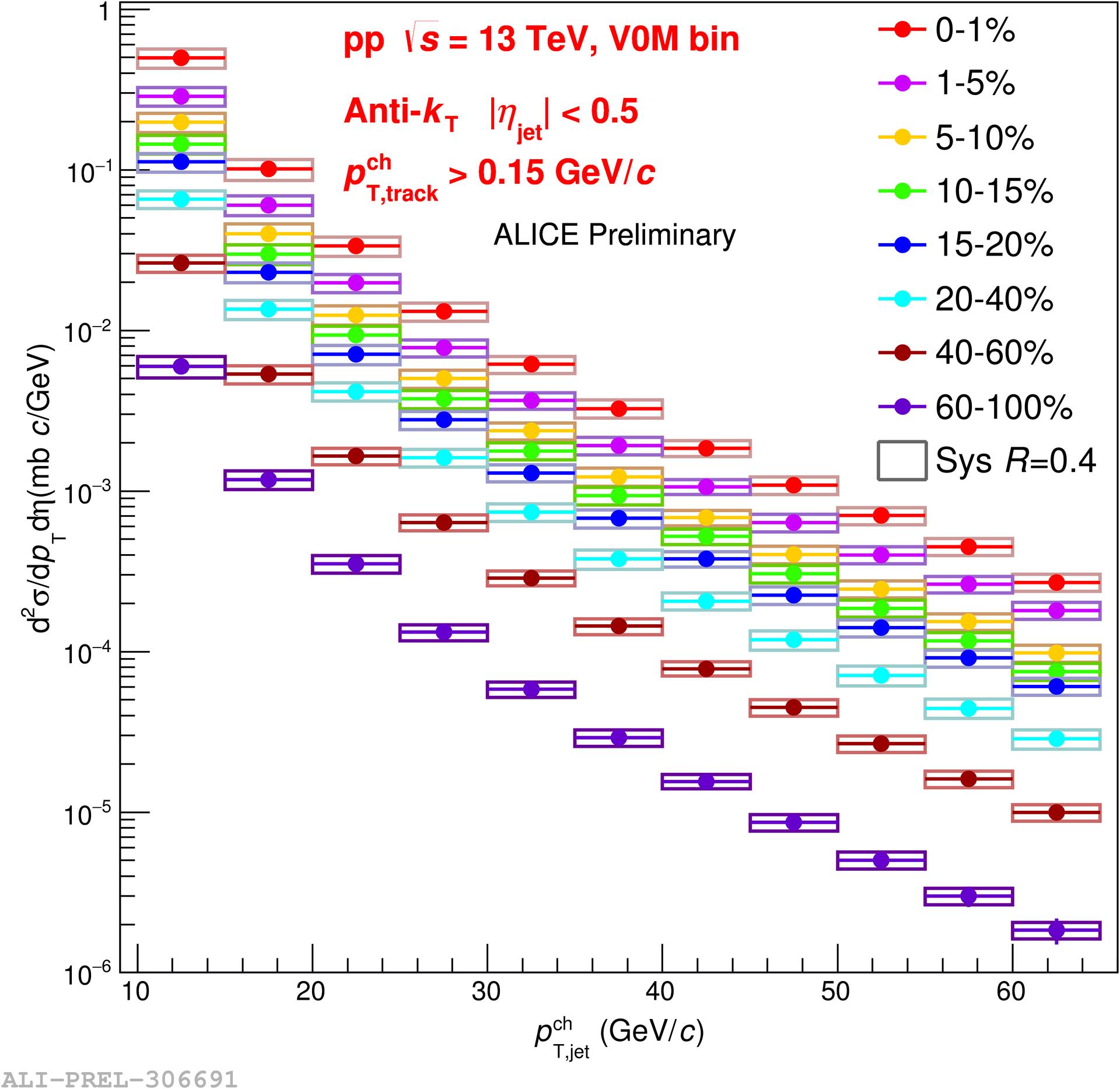}
    \includegraphics[width=0.3\textwidth]{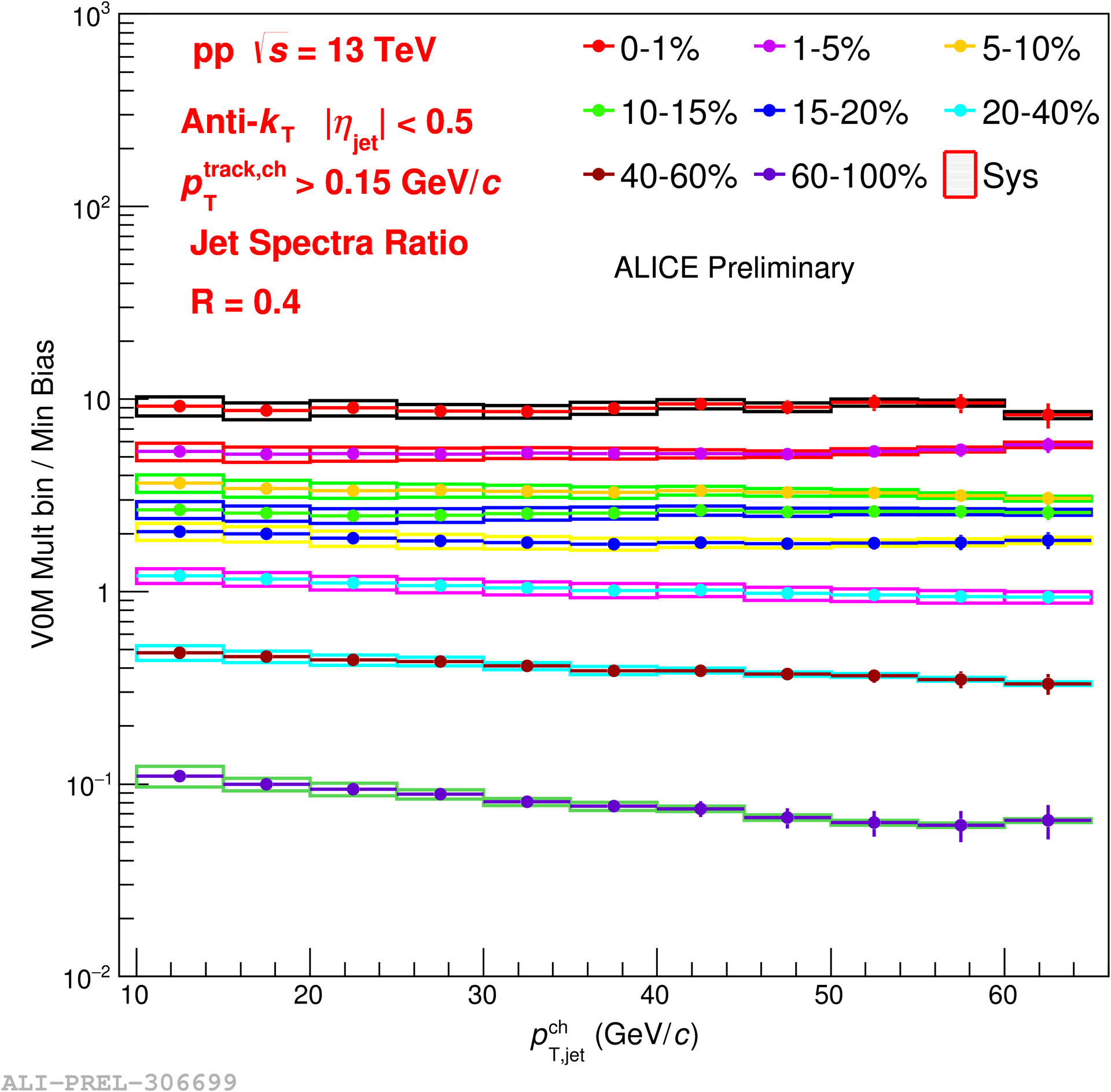}
    \includegraphics[width=0.3\textwidth]{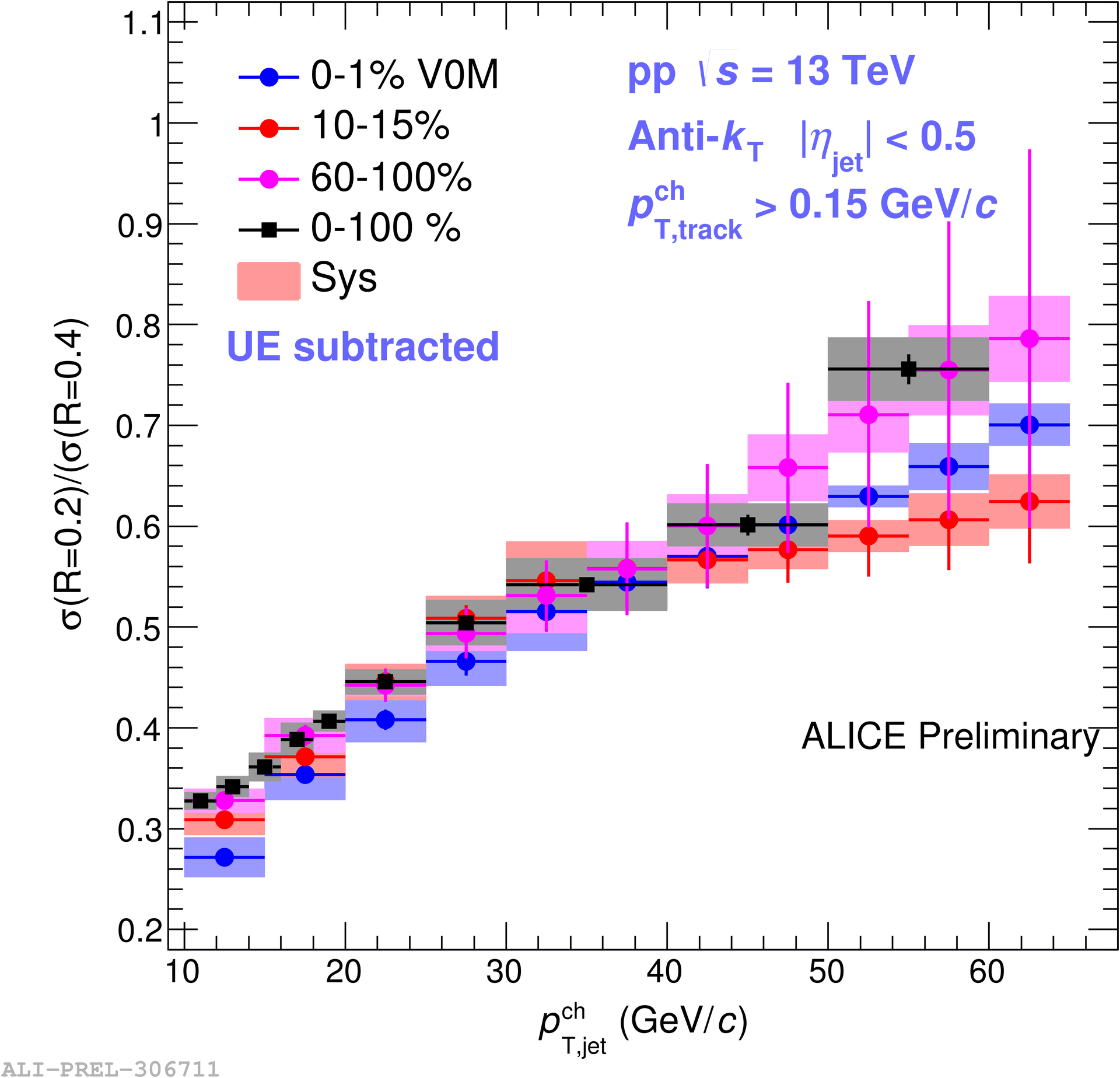}
    \caption{Multiplicity-dependence of track-based jet production: left: $p_{\rm{T}}$-differential cross section of track-based jet production in bins of the charged particle multiplicity. Middle: Ratio of the cross section in bins of multiplicity to the multiplicity-independent production cross section. Right: Cross section ratio of jet $R$=0.2/$R$=0.4 in three bins of multiplicity.}
    \label{fig:chargedJetsVsMult}
\end{figure}

The production of track-based jets has been measured with respect to the event activity expressed as the charged particle multiplicity measured with the VZERO detector. Fig.~\ref{fig:chargedJetsVsMult} left shows the $p_{\rm{T}}$-differential cross section of the production of track-based jets in bins of the multiplicity percentile. The ratio to the multiplicity-inclusive cross section is shown in the middle panel. A weak dependence on $p_{\rm{T}}$ can be observed, with a lower yield at high-$p_{\rm{T}}$ for jets in the lower multiplicity classes. The cross section ratio of $R$=0.2/$R$=0.4 jets is shown in the right panel for three characteristic multiplicity bins. An ordering with multiplicity can be observed, with the smallest cross section ratio in the highest multiplicity class.

\section{Measurement of the groomed momentum fraction $z_{g}$}

The groomed momentum fraction has been measured for fully reconstructed jets with $\rm{30}~\rm{GeV}/\it{c} \\ < \it{p}_{\rm{T}} < \rm{200}~\rm{GeV}/\it{c}$ and $0.2 < R < 0.5$. Jets were reclustered with the Cambridge/Aachen algorithm. As SoftDrop settings $z_{cut} = 0.1$ and $\beta = 0$ were used. No underlying event subtraction is applied as the grooming already removes the soft component. The $z_{\rm{g}}$ distributions were corrected back to particle level using a Bayesian unfolding in two dimensions. Results are shown for three $p_{\rm{T}}$ bins representing characteristic regions.

\begin{figure}
    \centering
    \includegraphics[width=0.3\textwidth]{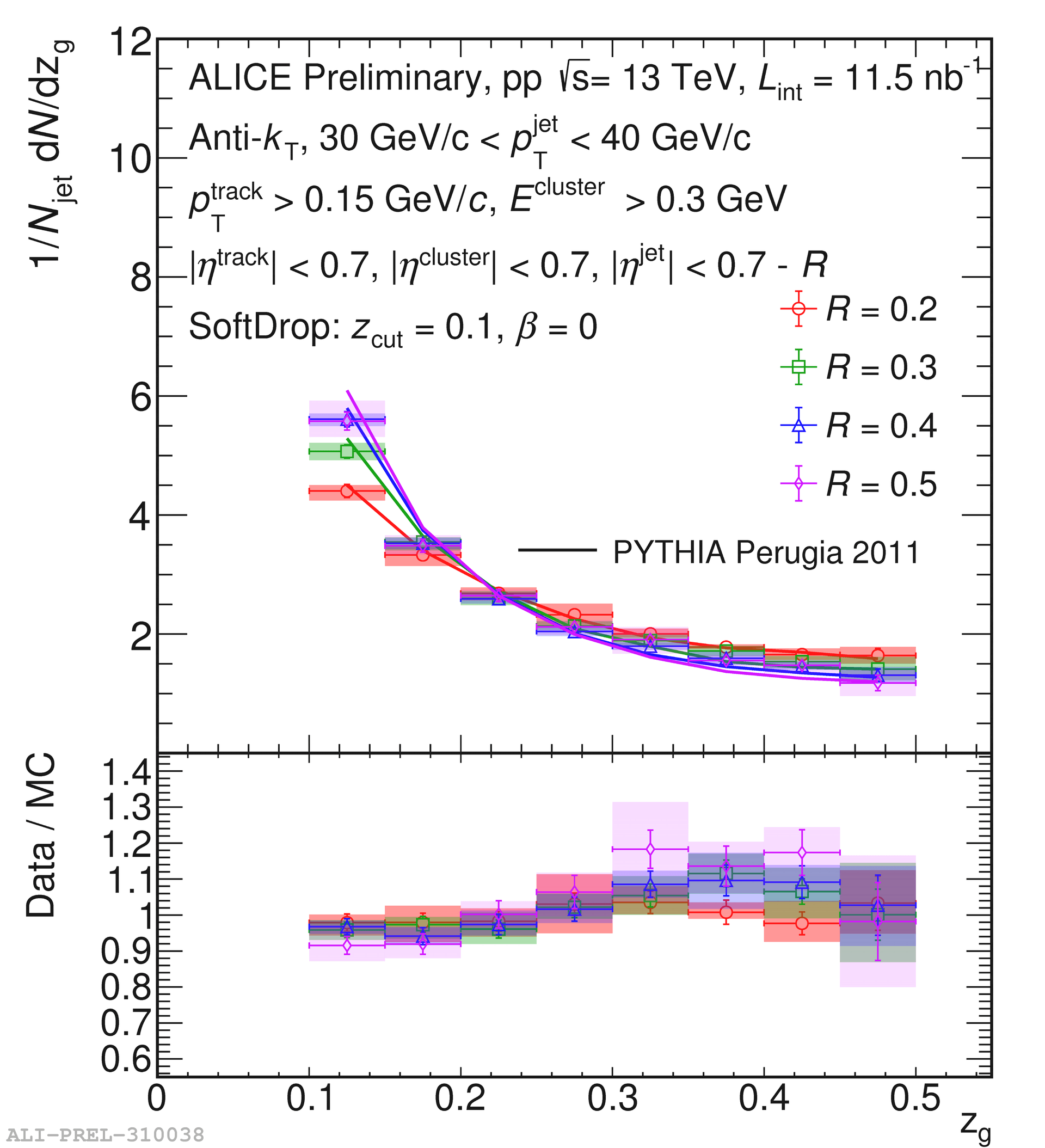}
    \includegraphics[width=0.3\textwidth]{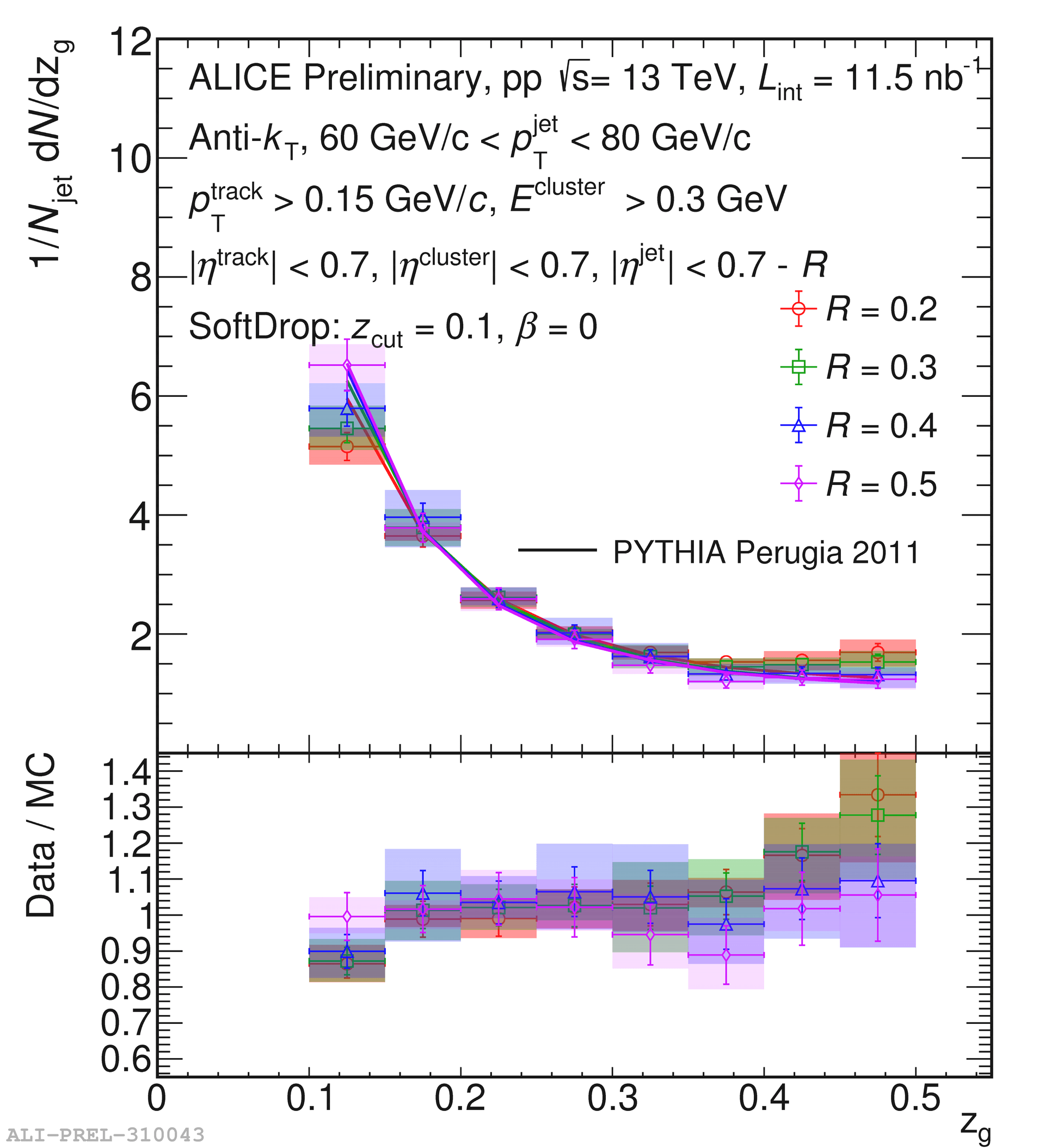}
    \includegraphics[width=0.3\textwidth]{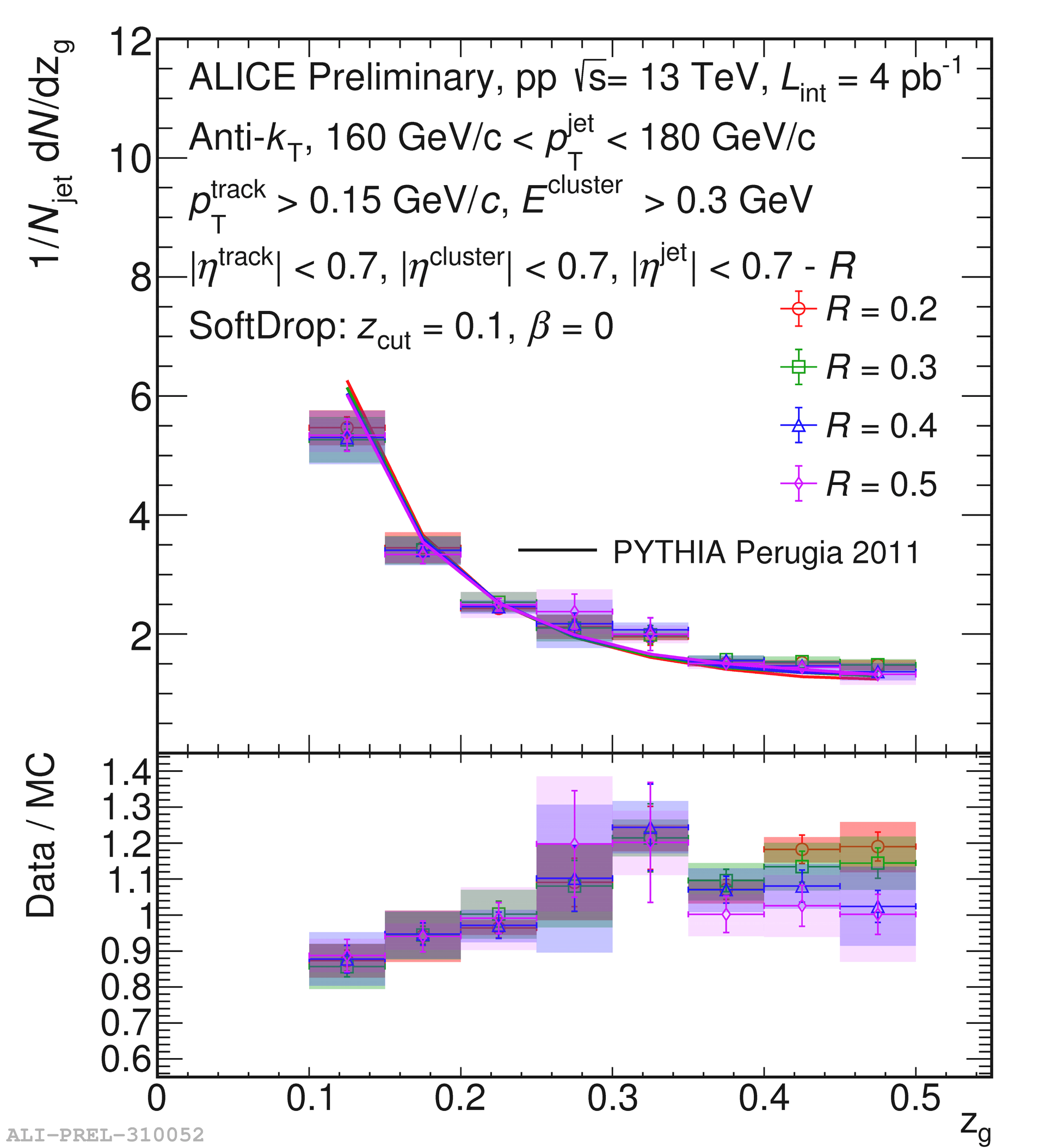}
    \caption{Dependence of the groomed momentum fraction $z_{\rm{g}}$ on the jet resolution parameter $R$ for jets with $\rm{30}~ \rm{GeV}/\it{c} < \it{p}_{\rm{T}} < \rm{40}~\rm{GeV}/\it{c}$ (left), $\rm{60}~\rm{GeV}/\it{c} < \it{p}_{\rm{T}} < \rm{80}~\rm{GeV}/\it{c}$ (middle) and $\rm{160}~\rm{GeV}/\it{c} < \it{p}_{\rm{T}} < \rm{180}~\rm{GeV}/\it{c}$ (right). Lines indicate the distributions obtained using the PYTHIA6 event generator with the Perugia2011 tune. Lower panels indicate the ratio of the data to PYTHIA.}
    \label{fig:zgvsRbinpt}
\end{figure}

Fig.~\ref{fig:zgvsRbinpt} shows the dependence of the $z_{\rm{g}}$ on the jet resolution parameter $R$ for  $\rm{30}~\rm{GeV}/\it{c} < \it{p}_{\rm{T}} < \rm{40}~\rm{GeV}/\it{c}$ (left), $\rm{60}~\rm{GeV}/\it{c} < \it{p}_{\rm{T}} < \rm{80}~\rm{GeV}/\it{c}$ (middle) and $\rm{160}~\rm{GeV}/\it{c} < \it{p}_{\rm{T}} < \rm{180}~\rm{GeV}/\it{c}$ (right). In the intermediate and high-$p_{\rm{T}}$ bin no dependence on the jet resolution parameter is observed. It can be understood as the dominant part of the jet energy is captured in the core of the jet and the large angle soft radiation plays only a minor role. For jets in the lowest $p_{\rm{T}}$ bin the shape of the $z_{\rm{g}}$ distribution differ with $R$. Jets with larger resolution parameter tend towards more asymmetric splitting, pointing to soft large angle radiation being recovered at larger jet resolution parameter. The separation of the distributions gradually reduces with increasing jet $p_{\rm{T}}$.

\begin{figure}
    \centering
    \includegraphics[width=0.3\textwidth]{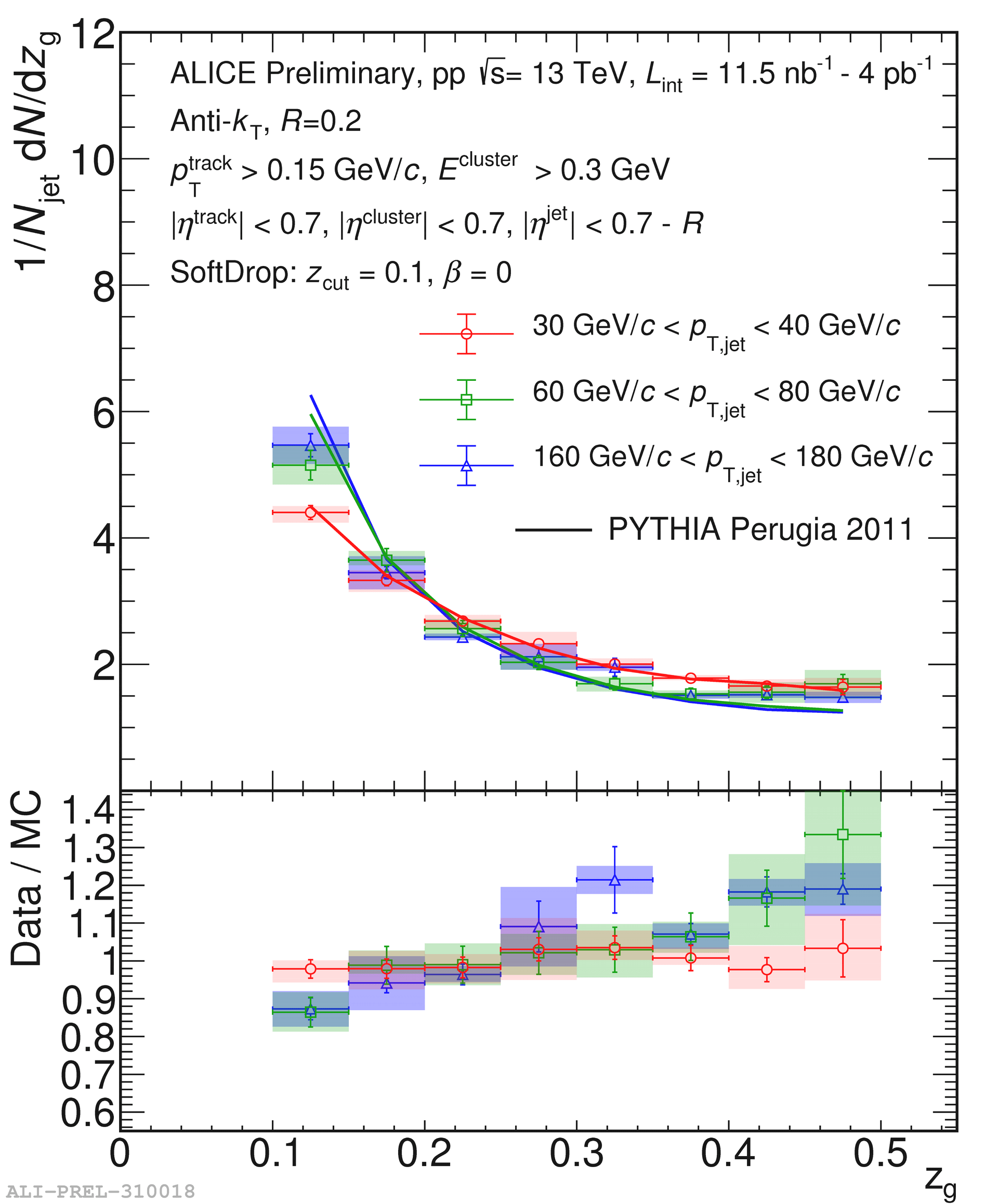}
    \includegraphics[width=0.3\textwidth]{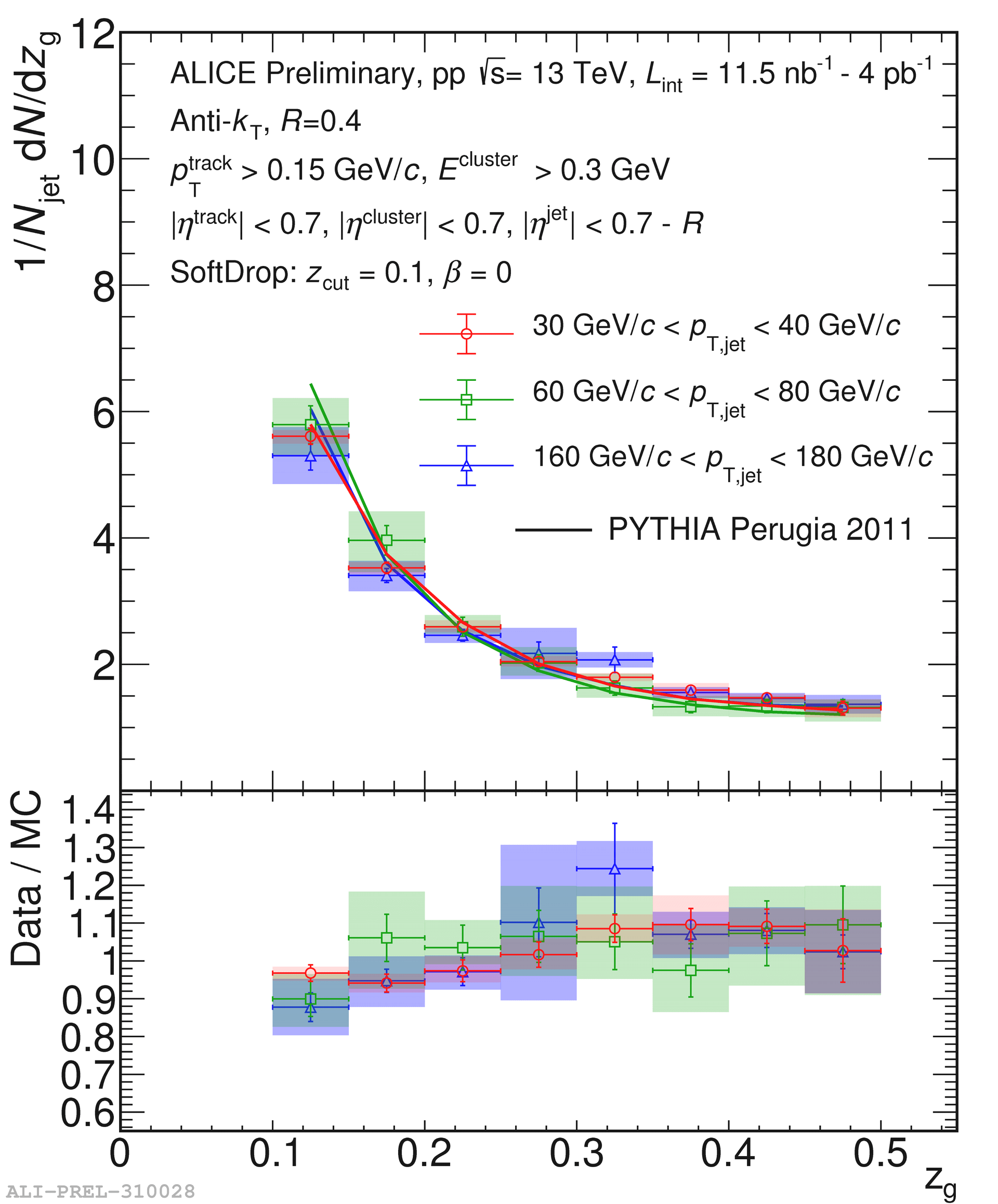}
    \caption{$p_{\rm{T}}$-dependence of the $z_{\rm{g}}$-distribution for jets with $R$=0.2 (left) and $R$=0.4 (right).Lines indicate the distributions obtained using the PYTHIA6 event generator with the Perugia2011 tune. Lower panels indicate the ratio of the data to PYTHIA.}
    \label{fig:zgvsptbinR}
\end{figure}

Looking at the dependence of the $z_{\rm{g}}$ distribution on the jet $p_{\rm{T}}$, shown in Fig~\ref{fig:zgvsptbinR} for $R$=0.2 and $R$=0.4, no dependence is observed for the larger jet resolution parameter while a moderate dependence with more symmetric splitting at lower jet $p_{\rm{T}}$ has been observed for the smaller jet resolution parameter.

The measurements are compared to calculations imposing the PYTHIA6 event generator using the Perugia2011 \cite{Skands:2010ak} tune. PYTHIA describes well both the dependency of the $z_{\rm{g}}$ on the jet $p_{\rm{T}}$ and the jet resolution parameter.

\section{Summary}
Jet substructure observables were measured in a wide range of jet resolution parameter and jet $p_{\rm{T}}$. Cross section ratios are in good agreement with PYTHIA and POWHEG+PYTHIA. A weak dependence of the jet production with the event activity has been observed. Except for the lowest $p_{\rm{T}}$ bin the $z_{\rm{g}}$ distribution is independent of $p_{\rm{T}}$, while at low $p_{\rm{T}}$ a dependence on the jet resolution parameter has been observed. A more conclusive picture can be obtained by extending the SoftDrop measurement to more shapes, among them $R_{\rm{g}}$ and $n_{\rm{SD}}$.

\bibliographystyle{JHEP}
\bibliography{biblio}{}

\end{document}